%% file: main.tex
\documentclass[twocolumn,switch]{article} % Method A for two-column 
\input{_format}
\usepackage{preprint}

\usepackage[numbers,square]{natbib}
\newcommand{\muc}{{\it M-UC}\xspace}
\usepackage[utf8]{inputenc}	% allow utf-8 input
\usepackage[T1]{fontenc}	% use 8-bit T1 fonts
\usepackage{csquotes}
\usepackage{xcolor}		% colors for hyperlinks
\usepackage[colorlinks = true,
            linkcolor = purple,
            urlcolor  = blue,
            citecolor = cyan,
            anchorcolor = black]{hyperref}	% Color links to references, figures, etc.
\usepackage{booktabs} 		% professional-quality tables
\usepackage{nicefrac}		% compact symbols for 1/2, etc.
\usepackage{microtype}		% microtypography
\usepackage{lineno}		% Line numbers
\usepackage{float}			% Allows for figures within multicol
\usepackage{multicol}		% Multiple columns (Method B)

\usepackage{lipsum}		%  Filler text
\usepackage{orcidlink}

\usepackage{newfloat}
\DeclareFloatingEnvironment[name={Supplementary Figure}]{suppfigure}
\usepackage{sidecap}
\usepackage{longtable}
\usepackage{array}
\usepackage{tabularx}
\sidecaptionvpos{figure}{c}
\usepackage{titlesec}
\titlespacing\section{0pt}{12pt plus 3pt minus 3pt}{1pt plus 1pt minus 1pt}
\titlespacing\subsection{0pt}{10pt plus 3pt minus 3pt}{1pt plus 1pt minus 1pt}
\titlespacing\subsubsection{0pt}{8pt plus 3pt minus 3pt}{1pt plus 1pt minus 1pt}

\usepackage{circledsteps}
\usepackage{tikz,xcolor,hyperref}
\usepackage{balance}
\title{Towards a Value-Complemented Framework for Enabling Human Monitoring in Cyber-Physical Systems}

\usepackage{eso-pic} % For adding custom text to specific pages

\AddToShipoutPictureBG*{%
  \ifnum\value{page}=1
    \put(0,20){%
      \makebox[\paperwidth]{\hfill \small\tt{*correspondence: Zoe.Pfister@uibk.ac.at} \hfill}%
    }%
  \fi
}
\usepackage{authblk}

\author[1\thanks{\tt{Zoe.Pfister@uibk.ac.at}}]{Zoe Pfister \orcidlink{0009-0009-2882-5059}}
\author[1]{Michael Vierhauser \orcidlink{0000-0003-2672-9230}}
\author[2,3]{Rebekka Wohlrab \orcidlink{0000-0002-5449-7900}}
\author[1]{Ruth Breu}

\affil[1]{University of Innsbruck\\Department of Computer Science\\Technikerstra\ss e 21a\\6020 Innsbruck\\Austria}
\affil[2]{Chalmers University of Technology and University of Gothenburg\\Sweden}
\affil[3]{Carnegie Mellon University\\Pittsburgh\\15213\\USA}

\begin{document}

\twocolumn[ % Method A for two-column formatting
  \begin{@twocolumnfalse} % Method A for two-column formatting
  
\maketitle

\input{sec_00_abstract}
\vspace{0.35cm}

  \end{@twocolumnfalse} % Method A for two-column formatting
] % Method A for two-column formatting

\input{sec_01_introduction}
\input{sec_02_background}
\input{sec_03_approach}
\input{sec_04_roadmap}
\input{sec_05_conclusion}

%%%%%%%%%%%%%%   Bibliography   %%%%%%%%%%%%%%
\normalsize

\bibliography{references} 
\balance

\end{document}

%% file: _format.tex
\usepackage[T1]{fontenc}
\usepackage{booktabs} % For formal tables
\usepackage[hyphens]{url}
\usepackage[table,xcdraw]{xcolor}
\usepackage{multirow}
\usepackage{rotating}
\usepackage{array}
\usepackage{color}
\usepackage{wasysym}
\usepackage{verbatim}
\usepackage{bigstrut}
\usepackage{multirow}
\usepackage{setspace}
\usepackage{tabularx}
\usepackage{url}
\usepackage{xspace}
\usepackage{pifont}
\usepackage{subcaption}
\usepackage[font=footnotesize]{caption}%,labelfont=bf]{caption}
\usepackage{array,booktabs}
\usepackage{soul}
\usepackage{vcell}
\usepackage{pifont}
\usepackage{afterpage}
\usepackage{rotating}
\usepackage{wrapfig}
\usepackage{caption}
\usepackage{graphicx}
\usepackage{caption}
\usepackage{enumitem}
\usepackage{lipsum}
\usepackage{balance}
\usepackage[ruled,vlined]{algorithm2e}
\usepackage{algpseudocode}
\usepackage{tikz}
\newcolumntype{L}[1]{>{\raggedright\let\newline\\\arraybackslash}p{#1}} 
\newcolumntype{C}[1]{>{\centering\let\newline\\\arraybackslash}p{#1}} 
\newcolumntype{M}[1]{>{\centering\arraybackslash}m{#1}}
\newcolumntype{R}[1]{>{\raggedleft\let\newline\\\arraybackslash}p{#1}} 
\newcommand{\citesec}[1]{Section~\ref{sec:#1}}
\newcommand{\citefig}[1]{Fig.~\ref{fig:#1}}

\newcommand{\etal}[1][]{%
\ifthenelse{\equal{#1}{}}{\emph{et~al.}\xspace}{\emph{et~al.}~\cite{#1}\xspace}%
}

\definecolor{alizarin}{rgb}{0.82, 0.1, 0.26}
\newcommand{\bullitem}[1]{\noindent\textbf{#1}}

%% file: sec_00_abstract.tex
\begin{abstract}
[\textbf{Context and Motivation}]:
Cyber-Physical Systems (CPS) have become relevant in a wide variety of different domains, integrating hardware and software, often operating in an emerging and uncertain environment where human actors actively or passively engage with the CPS. To ensure correct and safe operation, and  self-adaptation, monitors are used for collecting and analyzing diverse runtime information.
[\textbf{Problem}]:
However, monitoring humans at runtime, collecting potentially sensitive information about their actions and behavior, comes with significant ramifications that can severely hamper the successful integration of human-machine collaboration. Requirements engineering (RE) activities must integrate diverse human values, including Privacy, Security, and Self-Direction during system design, to avoid involuntary data sharing or misuse.
%Furthermore, to ensure acceptance of a system that analyzes and processes sensitive human-related data, trustworthiness needs to be established already when defining core requirements, particularly for any involved human-monitoring components. 
[\textbf{Principal Ideas}]:
In this research preview, we focus on the importance of incorporating these aspects in the RE lifecycle of eliciting and  creating runtime monitors.
% Particularly, informed by research on human values and value tactics, we created a conceptual framework for creating runtime monitors that actively include human-related properties.
[\textbf{Contribution}]:
We derived an initial conceptual framework, building on the value taxonomy introduced by Schwartz and human value integrated Software Engineering by Whittle, further leveraging the concept of value tactics. The goal is to tie functional and non-functional monitoring requirements to human values and establish traceability between values, requirements, and actors. Based on this, we lay out a research roadmap guiding our ongoing work in this area.
\end{abstract}

%% file: sec_01_introduction.tex
\section{Introduction}
\label{sec:intro}
% motivation with shop floor / self driving cards

%\zp{UC example excel: \href{https://uibkacat-my.sharepoint.com/:x:/g/personal/zoe_pfister_uibk_ac_at/EZMFDkVHlqNLsB9-myleJgMBXRRsXLDdOEclEAfMKdDhlw?e=vCMj0Q}{Use-Case Vignettes}}

%\mv{context and motivation:}
Cyber-Physical Systems (CPS) have become increasingly prevalent in our society, including, for example, autonomous vehicles or drones.
%The use cases of these systems range from autonomous transportation and environmental monitoring to search-and-rescue missions. 
As part of shop-floor automation, Cyber-Physical Production Systems integrate robots to perform tasks like automated material transport~\cite{theunissen2018smart}.
Typically, CPS exhibit tight integration of hardware and software, combining virtual and physical spaces~\cite{broy2013engineering}. Moreover, CPS often operate in emerging, uncertain environments alongside human stakeholders, such as shop floor workers, drone operators, or pedestrians.

% Monitoring CPS is critical for ensuring safe operation and for enabling self-adaptive behavior, with monitors deployed that collect runtime data and perform constraint checks at runtime.
To ensure safe operation and enable self-adaptive behavior, it is critical to monitor CPS through deployed monitors that collect data and perform constraint checks at runtime.
%,zheng2016efficient}. 
In recent years, paradigms such as Human-Machine Teaming have emerged, where humans are more closely and actively collaborating with CPS.
Current monitoring solutions often consider safety concerns in conjunction with human-machine interaction~\cite{tan2011triple}.
To ensure safety in scenarios with tight human-machine interaction, mechanisms need to be in place not only for the monitoring of the system and collecting properties at runtime, but also of human actors with properties such as location, posture, or cognitive load~\cite{mukhopadhyay2014wearable}. Collecting and analyzing information about a human actor engaged with a CPS can provide additional benefits, for example, depending on the position, movement, and breath data of a shop-floor worker, the system may decide to send emergency personnel to investigate if an accident has occurred~\cite{nwakanma2021Detection}.

%In the context of human-machine collaboration and self-adaptation, monitoring data about the human operator can be leveraged by the system to notify emergency services in case it detects worker injury, prompt the user for inputs, or trigger certain actions when humans are detected during a search-and-rescue mission.

% For example, depending on the current environmental context and the state of a human operator, the system may decide to reduce the information provided to the operator and operate more autonomously, with the aim to reduce their cognitive load~\cite{cleland-huang2023HumanMachine}.
%\mv{Question and Problems:}
However, monitoring humans at runtime, and collecting information about their actions and behavior, comes with significant ramifications that can severely hinder the successful integration of human-machine collaboration in a CPS. Recently, Whittle~\etal[whittle2021Case] argued that \emph{\enquote{human values are heavily underrepresented in SE methods}}.
This is particularly true when sensitive data about humans is collected that can potentially be used in ways it was not intended. For example, Amazon has made headlines for monitoring the workers of their warehouses to track and assess worker performance\footnote{\url{https://tinyurl.com/amazon-monitoring}, accessed Nov 1, 2024.}.
This example reinforces that aspects such as \emph{Privacy} -- and human values in general -- are a major concern, as sensitive data might be stored, shared, or misused. In addition, \textit{ethical aspects} regarding the responsibility of actions must be investigated. 
Questions like, ``who is responsible for an action within a CPS with tight human-machine collaboration?'' and ``when should the system be allowed to act fully autonomously?'' must be considered during design and implementation of the system.
Besides privacy and ethics, it is crucial to consider the \textit{technical aspects} of human monitoring in CPS. Instead of ``simply'' collecting data from the system through sensors, human monitoring often requires additional hardware, such as cameras or wearables, further adding to the complexity.  %For example, monitoring the position of shop-floor workers can be achieved through LIDAR sensors~\cite{nwakanma2021Detection}. 

In this research preview, we focus on the importance of incorporating human values in the RE process, addressing challenges and shortcomings of existing approaches (\citesec{background}),  proposing an initial conceptual Value-Complemented Framework (\citesec{approach}), and discussing our research roadmap (\citesec{roadmap}).

%% file: sec_02_background.tex
\section{Background \& Motivating Example}
\label{sec:background}
% also relwork
% \mv{current state of the arg  - problems and challenges - why HMT is needed for CPS!}
%https://www.mdpi.com/1424-8220/23/13/6054
% \mv{
% challenges:
% -autonomy
% -heterogeneity
% complexity
% ...
% }
% ==> 
Runtime monitoring has been an active research area for many years, ranging from model-based approaches and self-adaptive systems, to formal verification and  model-checking of monitored properties~\cite{vierhauser2016requirements}. 
However, hardly any of these approaches has considered human actors, let alone human values or ethical considerations, when collecting potentially sensitive runtime data.
Human aspects are, to some extent, addressed in the Human-Computer Interaction (HCI) community, e.g., as part of supporting human operators in complex, safety-critical environments~\cite{schmid20}, but not directly in the context of runtime monitoring.
On the other hand, integrating human values into software engineering has received increased attention in recent years. To do so, we first need a taxonomy of potential human values.
One of the most prominent human value theories is Schwartz's theory~\cite{schwartz1992Universals}, which includes a list of 10 core universal values and various sub-values, further showcasing which values complement or oppose each other.
For example, the universal value \textit{Security} contains sub-values such as national security, which is opposed to the universal value \textit{Self-Direction} that includes privacy.
%In contrast, values like freedom and broad-mindedness complement each other.
Focusing on SE activities, Whittle \etal[whittle2021Case] have used the value taxonomy of Schwartz to enhance the requirements engineering process. 
%They argue that human values \enquote{should be considered as first-class citizens in software development}.
%In an experimental software project in collaboration with \textit{Clasp} -- a project with the aim to reduce social anxiety in autistic adults --, they defined relevant values for each stakeholder in the project. % iterative process
%Stakeholders included, for example, users of the final software, the product owners, and the project team itself.
% The values are context specific to the individual project. % value portraits
From a set of identified values, they then extracted new requirements.
% they don't filter it by all the values, but by a selection of four overarching things (openness to change, self-enhancement, self-transcendence, conservation
During an experimental software project, they found that the values defined as part of their process capture the \textit{why} of requirements engineering, complementing the traditional view of \textit{what} and \textit{how} of functional and non-functional requirements respectively.

Connecting these aspects, we have previously presented an approach to integrate human values into the RE process~\cite{wohlrab2024Supporting}.
In our \enquote{value-aware requirements engineering} framework, original system goals are enriched with relevant values and sub-values based on the taxonomy of Schwartz~\cite{schwartz1992Universals}.
%After the initial definition, values are prioritized, and consolidated into personas.
As a next step, so-called \enquote{value tactics} are selected for the values and sub-values. Value tactics are decisions that directly address a particular sub-value and can be used to derive functional and non-functional requirements.
An example value tactic for the value-sub-value pair \textit{benevolence-honesty} is to \textit{anonymize transactions}. These value tactics then enable system engineers to elicit detailed requirements, e.g., \textit{anonymize transactions} could lead to specific data protection requirements.

However, the individual aspects of system use cases, human values, and ultimately the resulting requirements are still somewhat disconnected, and streamlining these activities in a coherent RE process remains an open challenge. Particularly in the context of (human) runtime monitoring, this aspect is crucial as different values and their resulting requirements can have an immediate impact on, for example, what data can/must be collected and can be stored or must be anonymized/deleted.
To motivate the challenges and benefits of human-monitoring at runtime in CPS, we present an example Use Case (cf. Table~\ref{tab:motivating-example}) that promotes increasing worker safety in a shop-floor environment.
In the Use Case snippet, we define that the system must detect when a shop-floor worker enters a restricted or dangerous area (e.g., areas with autonomous robots). % which may lead to injuries
Through continuous monitoring of a worker's position, the system must notify the worker if they move beyond the specified boundary. Additionally, emergency personnel must be alerted if the worker does not respond within a certain amount of time.
A system implementing this use case can thus improve the safety of workers, but comes with privacy concerns.
Ultimately, the values \textit{security} (protecting individuals from threats) and \textit{independence} (personal privacy) need to be traded off against each other.
This paper proposes a framework to do that and arrive at a set of value-complemented monitoring requirements.

\input{shop-floor-uc-table}

% \zp{describe motivating example and human values that arise from that in textual form and show the tables later. Bring amazon privacy examples (see verge article).}
% \zp{in a shop floor there are workers, robots, dangerous areas requiring safety aspects. }

% HMT examples: emergency response https://doi.org/10.3390/app11083662
% 
% Bashar: https://dl.acm.org/doi/abs/10.1145/3640310.3674095

% Requirements reflection: requirements as runtime entities
% N Bencomo, J Whittle, P Sawyer, A Finkelstein

% how does whittles process look like?

% wohlrab
% similar approach based on value derivation and persona creation. They then create 'value-tactics' which lead to the definition of requirements. ““value-aware requirements engineering”” 

%% file: shop-floor-uc-table.tex
\begin{table}[hbt]
\footnotesize
\caption{Motivating Use Case Example.}
\renewcommand*{\arraystretch}{1.22}
\label{tab:motivating-example}
\begin{tabularx}{\columnwidth}{L{2.75cm}X}

\toprule
% Use Case related to & Detect Worker Emergency \\
Name & Detect worker entering hazardous area. \\
Description & The system tracks shop-floor workers and notifies them if they enter a restricted or dangerous area. \\
Pre-Conditions & (1) The shop floor worker is equipped with location tracking devices OR the shop floor is equipped with LiDAR to track workers. 

(2) The worker enters a restricted area. \\
Trigger & An administrator starts the tracking software OR The software continuously tracks workers in an area. \\
Success End Condition & The worker is notified when entering a restricted area. If they do not leave within a specified  time, additional emergency personnel are notified. \\

% Error Situations & The location tracking is defective. \\
Failure End Condition & The worker enters a restricted area but is not notified.\\
\midrule
Monitoring Targets & Shop-Floor Worker, Robot \\
Monitored Properties & movement data / worker position / tracking device OR LiDAR\\

\emph{Additional Actors} & Location Tracking Devices, Emergency Personnel, \enquote{Others}, Floor Manager / Shift Supervisor\\

\bottomrule
\end{tabularx}
\renewcommand*{\arraystretch}{1}
\end{table}

%% file: sec_03_approach.tex
\section{Enabling Value-Complemented Human Monitoring}
\label{sec:approach}
% value lead that confirms the values and approves them through signatures from stakeholders
In our proposed approach, the goal is to actively incorporate human values as first-class citizens in the early stages of the requirements engineering process, guiding subsequent elicitation of requirements, and ultimately system design. 

\bullitem{Framework Overview:}
We achieve this by first identifying relevant values and corresponding stakeholders, followed by mapping them to relevant monitoring properties (i.e., what data we want to collect at runtime) within the context of a specific \emph{Monitoring Use Case (\muc)}.
\citefig{value-based-framework} provides a high-level overview of our framework, guiding the value complemented monitoring RE process.
% Commonly in requirements engineering, we directly derive both functional and non-functional requirements based on a system usecase. However, this results in human-values being ignored in the process.
In a first step \Circled{1},  we identify the stakeholders \textit{who} will engage with the feature described in the System Use Case, and would be impacted by monitoring functionality, and create respective personas~\cite{schneidewind2012How}. 
Based on these, we then derive a specific \emph{Monitoring Use Case} \Circled{2}, a  subset of a traditional system use case~\cite{yue2013facilitating}, specifically focusing on monitoring-related elements, components, and stakeholders. %These parts are then further augmented with information concerning the potentially relevant information (properties) that should be collected at runtime.

\begin{figure*}[t!]
    \centering
    \includegraphics[width=.95\textwidth]{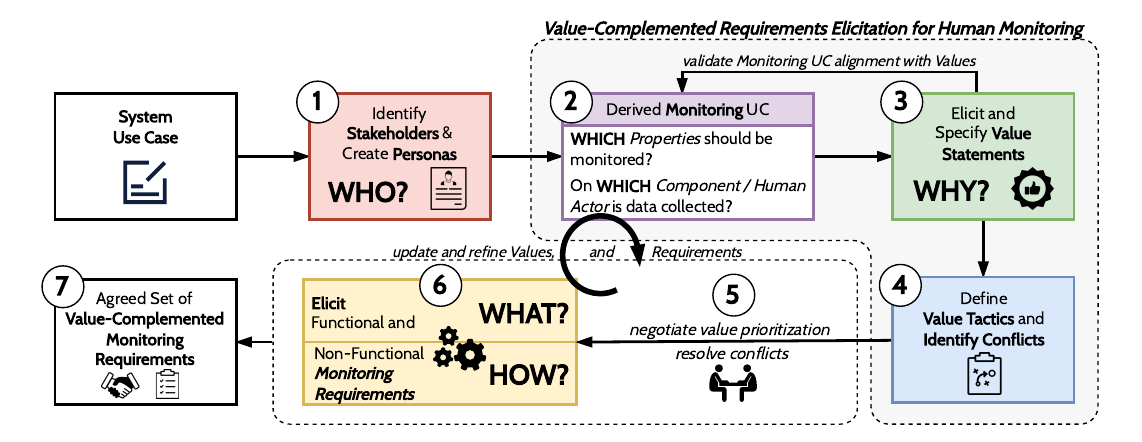}
    \vspace{-10pt}
    \caption{Conceptual value-complemented Framework.}
    \label{fig:value-based-framework}
\end{figure*}

The \muc is then annotated with information regarding \textit{which} properties should be monitored and from \textit{which} component or human actor the data is collected. 
Following Whittle~\etal's~\cite{whittle2021Case} work on human values,  we then identify relevant values \Circled{3} for each of the identified personas and thus their respective stakeholders.
We hereby leverage the taxonomy of Schwartz~\cite{schwartz1992Universals} that categorizes values into Self-Enhancement, Conservation, Openness to Change, and Self-Transcendence. Depending on the stakeholders and monitoring properties involved in a \muc, the defined values can vary significantly~\cite{Spiekermann2023Value}. %p100
These identified values add the dimension of \textit{why} stakeholders want to see a feature implemented, and more crucially, why it should be implemented in a certain way (e.g., by obfuscating user information).
Based on these persona-value pairs, we then define \textit{value tactics}~\cite{wohlrab2024Supporting} in step \Circled{4}, which serve as a link between human-values and functional and non-functional requirements addressing specific quality attributes or suggesting mechanisms to consider in later requirement elicitation.

In step \Circled{4}, we also identify which value tactics oppose or complement each other, serving as value-related input for value prioritization.
Once the initial set of value tactics and their conflicts have been resolved \Circled{5}, the process continues, focusing on eliciting detailed monitoring requirements \Circled{6}, i.e., requirements that are related to the monitoring of actors.
The results of the value tactic and conflict resolution steps serve as input for the monitoring requirements elicitation phase of the overall system. 
The ``traditional'' RE process~\cite{pohl2010requirements}, is thereby  enriched by the notion of the previously identified monitoring properties, personas, their values, and resulting value tactics.
As shown in Figure~\ref{fig:value-based-framework}, each of the described steps are repeated until finally, all affected stakeholders agree upon a set of requirements for the monitoring UC \Circled{7}.
Based on these artifacts, system engineers are then able to design and implement the monitoring infrastructure.

%%% pointer to (5) is still missing

%\subsection{Applying the Process}
\bullitem{Applying the Framework:}
As a first proof-of-concept illustrating our approach and demonstrating it for a concrete scenario, we created an example \muc, personas, values, value tactics, and monitoring requirements, for the aforementioned Use Case (cf. Table~\ref{tab:motivating-example}).
More specifically, the UC revolves around detecting when a shop-floor worker enters a potentially dangerous and restricted area, e.g., where autonomous manufacturing robots are operating, and notifying them to leave said area\footnote{Due to space limitations, we only present a subset of values and tactics in this paper. All created artifacts are available in the supplemental material (\url{https://github.com/Ethical-Human-Machine-Interaction/vcmf}.}.
First, we identified the key stakeholders of the system use case and developed their personas. In our example, we found three key stakeholders: (a) Shop-Floor Workers, (b) Factory Managers / Supervisors, and (c) stakeholders that fall in the ``Others'' category (e.g., personnel without access to dangerous areas).
Based on the System Use Case, we then derived a concrete \muc that further details how monitoring is involved in the process, and particularly, how monitoring involves/affects the human stakeholders.
Next, we augment our \muc with additional information cornering \textit{which} properties should be monitored from \textit{which} components or human actors.
In our shop-floor scenario, this, for example, includes monitoring the position of workers, potentially with an IoT device that is integrated in protective work clothing, or a LiDAR scanner installed at relevant locations inside the factory. Additionally, this requires the collection of non-human-related information, such as the position of robots workers may collide with.

%For each of these stakeholder groups, we then created dedicated personas to gain a better understanding of \textit{who} embodies which values.
We then defined the potential values of each stakeholder group and organized them according to  Schwartz'~\cite{schwartz1992Universals} four value categories. % important to use a taxonomy to gain a better understanding of what needs to be thought of.
An example \textit{Self-Enhancement} value definition of a shop-floor worker for our \muc is ``I want to be able to withdraw my consent of monitoring, even after agreeing to data collection''.
On the other hand, a factory manager's \textit{Self-Enhancement} values may include ``I am able to hold employees accountable for their actions when human-caused accidents occur.''. After defining a set of values for each persona, we defined value tactics (VTs). 
Revisiting the examples from above, one VT is related to \enquote{\textit{allow opt-out of tracking}} and another to \enquote{\textit{track which shop-floor workers enter dangerous areas}}.
Afterward, we consolidated all VTs and identified potential conflicts. 
The two example VTs also exemplify conflicting values which need to be resolved before the final requirements elicitation phase. We argue that, to resolve conflicts, the individual values defined by the stakeholders must be prioritized. The IEEE 7000\textsuperscript{\texttrademark} standard~\cite{noauthor2021ieee} provides criteria to prioritize the core values of a System of Interest, which serves as a basis in this process. 
%Ultimately, the executive leaders assume accountability through their signature and publication of the final value prioritization~\cite{Spiekermann2023Value}.
We realize that value prioritization is an important issue, which is why we aim to develop processes that help stakeholders prioritize values in future work.
% \zp{value negotiation is an open issue and not solved yet}
% TODO: VALUE PRIORIZATION, value based WIN WIN

When eliciting the final requirements, it is crucial to specify values that are related, providing clear traces throughout the entire process.
For example, shop-floor workers not being allowed to opt out of tracking because of safety concerns is related to the   monitoring requirement  \enquote{\textit{When a user enters the bounds of a dangerous area, the system shall always notify the user by audiovisual notifications.}} % \zp{add monitoring specific usecase}
Notice that the wording of the requirement specifies that a user will \textit{always} be notified, meaning it is not possible to opt out of position tracking or disabling notifications by a shop-floor worker. 
Naturally, this would also lead to a discussion about tracking constraints, such as the frequency of measurements.
Such constraints can trigger additional iteration cycles to reassess and potentially revise the \muc, value statements, or value tactics. 
% \zp{add what exactely is the moniotring requirement - did in 3.1}
% \mv{maybe some concluding/discussion remarks...  Following this initial version of the process we were able to... \zp{see comment}}
% \zp{we require constraints}
% \zp{process steps. WHAT WHY HOW}

% schwartz value framework
% Actors / stakeholders / personas of CPS (i.e., shop floor)
% Privacy tactics: \url{https://doi.org/10.1109/SPW.2016.23}
% tactics (value based engineering, value tactics, constraints to these tactics)
% -> monitoring needs
% -> conflicts (i.e., privacy, ethics, .... based on schwartz and related to actors) -> use tactics to fit specific needs -> make a matrix of that
% -> IEEE 7000-2001 (annex G) lists some ``typical ethical values for system design''

% “Human behavior is complex and may not always conform to the model’s theoretical framework. Actors may exhibit contradictory or unexpected behaviors, necessitating flexibility in applying the model and interpreting its results.” ([Alrimawi and Nuseibeh, 2024, p. 6])

% what requirements can we inherit from the motivating example, how can we extend it then?

%% file: sec_04_roadmap.tex
\section{Research Roadmap \& Discussion}
\label{sec:roadmap}
%==> 1) Value Validation (Michael)
%% how to validate nfr/ethics
%%% coverage - >traceabilty 
%% post-morten tracing

%==> 2) monitoring Tactics taxonomy/catalogue 
% creation of a catalog of values to use in conjunction with value taxonomies
% also link that with potential value tactics

Based on this initial conceptual framework and application example, we identified several areas where further research is necessary to successfully combine human monitoring and value-complemented requirements engineering.

\bullitem{I -- Continuous Value Validation:} One key aspect is the validation of both the \muc, and resulting functional and non-functional monitoring requirements.  This ensures that human-value and ethical considerations are embedded not only throughout the elicitation process, but also during system operation. 
Traceability hereby plays a pivotal role, both during initial development and throughout the lifecycle, as systems evolve. A dedicated Traceability Information Model defining the artifacts and their respective trace links can facilitate {(semi-)automated} validation checks to ensure that crucial ethical considerations and conflicting aspects are properly covered~\cite{mader2009getting}.
As part of this, we will explore processes to establish traceability between monitoring properties and requirements. Additionally, we will apply validation techniques to both requirements and trace links to ensure that, for example, value decisions do not supersede safety-critical requirements, e.g., leveraging the concept of safety performance indicators~\cite{koopman2020positive}. %for both safety and security-related properties.

\bullitem{II - Monitoring Value Tactics Taxonomy:} A second research direction focuses on developing a catalog of monitoring value tactics. After all, there are common patterns of concerns that stakeholders raise and monitoring-related decisions that are beneficial -- independently of the concrete system, context, and stakeholders. We plan to follow guidelines for taxonomy building in software engineering~\cite{usman2017taxonomies}, which includes exploring existing value tactics~\cite{wohlrab2024Supporting} and how they can be translated to monitoring requirements. %We expect that certain tactics will be specific to human monitoring, especially when it comes to safety and privacy concerns.
%The table we include in the supplemental material can serve as a starting point.
Empirical studies with companies will contribute to this direction so that we can explore and elicit typical tactics that practitioners deem relevant.

\bullitem{III - Ethics-Aware Requirements Elicitation:}
Third, we will continue to explore methods that aid in value-complemented RE, such as addressing ethical concerns. %Ethics is about the concepts of right and wrong and must be considered when designing and developing self-adaptive CPS, as they are systems that operate within our society and make decisions based on our expectations regarding morality and fairness~\cite{trentesaux2020Ethical}.
Trentesaux and Karnouskos~\cite{trentesaux2020Ethical} specify two main types of ethics relevant for CPS from a system engineering context: (1) \enquote{actors involved in the design and production of the CPS}, and (2) \enquote{the ethics of the CPS itself during use}. %Both are relevant during the requirements elicitation process.
One particularly important aspect here are conflict resolution strategies. In future work, we plan to establish a more sophisticated process that assists stakeholders and developers in prioritizing their specified values. As a starting point, we will leverage existing literature~\cite{noauthor2021ieee,Spiekermann2023Value} that utilize \textit{core values} for value prioritization criteria.
In addition, we aim to create concepts of what aspects should be considered when implementing ethics-aware CPS. This includes, for example, a process of adding phases to the requirements engineering process that specifically add information on how a CPS must act in high-risk situations. 

Finally, our ongoing work focuses on implementing the framework in a broader  real-world context, performing a comprehensive validation study. This includes exploring human-monitoring requirements for robotic and drone  applications, building on existing work on runtime monitoring, %~\cite{vierhauser2018monitoring} 
and extending the framework with runtime monitors and checks using IoT and sensor devices for collecting location data, or cameras capturing, e.g., the position of a human operator.

% where humans could be in danger (trentesaux2020)
% general definition of ethics: “Ethics, as a field of philosophy, engages in concepts of right and wrong. For intelligent autonomous CPS, ethical behaviors become relevant since these are expected to operate within society.” (Trentesaux and Karnouskos, 2020, p. 58)
% what is involved: “Two main types of ethics from a system engineering point of view are directly relevant i.e. the ethics of the actors involved in the design and production of the CPS and the ethics of the CPS itself during its use [33].” (Trentesaux and Karnouskos, 2020, p. 58)
% conflict resolution is made topic in spiekermann and the IEEE standard. what is right and wrong -> core values of an organization
% 2 sentences on the ethics of the CPS itself
% 
% ==> 3) Decision Making for req. elicitation / conflict resolution (Zoe)
%% safety
%% ethical decision making
%% ethics as a first class citizen-> in architecutre/requirements

%\mv{MV: TECHNICAL ASPECTS!!!!}

%implement, broanden scpe and validate..
%=> same for constraints/stuff that is monitored
%human-models@runtime (When/where) - local/central - different contexts

%% ethical runtime adaptation
%%holistic SE process support
%% similar to usability -> atam++ vor values
% review on methods to integrate values and ethics into software engineering
% look into different value taxonomies, not just Schwartz

% iterative process for discovery / design

%%%TRACEABILITY!

%% file: sec_05_conclusion.tex
\section{Conclusion}
\label{sec:conclusion}
In this research preview paper,  we present our  ideas  towards incorporating human values in the process of  eliciting and specifying requirements for  creating runtime monitors that collect human-related data.
Our initial conceptual framework for value-complemented Human-Monitoring defines key activities, from specifying monitoring use cases to applying value tactics for creating a set of agreed on functional and non-functional monitoring requirements. 

We further identified three key areas as part of our research roadmap, related to elicitation, validation, and negotiation of values and requirements. These require further in-depth research and analysis in the context of CPS to tie functional and non-functional monitoring requirements to human values.